\documentclass[sigconf]{acmart}

\usepackage{comment}
\usepackage{float}
\restylefloat{table}
\usepackage{tabularx}
\usepackage{stfloats}
\usepackage{placeins}

\AtBeginDocument{%
  \providecommand\BibTeX{{%
    \normalfont B\kern-0.5em{\scshape i\kern-0.25em b}\kern-0.8em\TeX}}}

\setcopyright{acmcopyright}
\copyrightyear{2018}
\acmYear{2018}
\acmDOI{XXXXXXX.XXXXXXX}

\acmConference[Conference acronym 'XX]{Make sure to enter the correct
  conference title from your rights confirmation emai}{June 03--05,
  2018}{Woodstock, NY}
\acmPrice{15.00}
\acmISBN{978-1-4503-XXXX-X/18/06}

\usepackage{packages}

\title{ChatGPT for Conversational Recommendation: Refining Recommendations by Reprompting with Feedback}

\begin{document}

\author{Kyle Dylan Spurlock}
\email{kyle.spurlock@louisville.edu}
\authornotemark[1]
\affiliation{%
  \institution{University of Louisville}
  \city{Louisville}
  \state{Kentucky}
  \country{USA}
}

\author{Cagla Acun}
\email{a0acun01@louisville.edu}
\affiliation{%
  \institution{University of Louisville}
  \city{Louisville}
  \state{Kentucky}
  \country{USA}
}

\author{Esin Saka}
\email{esinsaka@gmail.com}
\affiliation{%
  \institution{Microsoft}
  \city{Seattle}
  \state{Washington}
  \country{USA}
}

\author{Olfa Nasraoui}
\email{olfa.nasraoui@louisville.edu}
\affiliation{%
  \institution{University of Louisville}
  \city{Louisville}
  \state{Kentucky}
  \country{USA}
}

\renewcommand{\shortauthors}{Spurlock et al.}


\begin{abstract}
    Recommendation algorithms have been pivotal in handling the overwhelming volume of online content. However, these algorithms seldom consider direct user input, resulting in superficial interaction between them. Efforts have been made to include the user directly in the recommendation process through conversation, but these systems too have had limited interactivity. Recently, Large Language Models (LLMs) like ChatGPT have gained popularity due to their ease of use and their ability to adapt dynamically to various tasks while responding to feedback. In this paper, we investigate the effectiveness of ChatGPT as a top-n conversational recommendation system. We build a rigorous pipeline around ChatGPT to simulate how a user might realistically probe the model for recommendations: by first instructing and then reprompting with feedback to refine a set of recommendations. We further explore the effect of popularity bias in ChatGPT's recommendations, and compare its performance to baseline models. We find that reprompting ChatGPT with feedback is an effective strategy to improve recommendation relevancy, and that popularity bias can be mitigated through prompt engineering.
\end{abstract}

\ccsdesc[500]{Information systems~Recommender systems}
\ccsdesc[500]{Information systems~Retrieval effectiveness}
\ccsdesc[500]{Information systems~Language models}

\keywords{Recommender Systems, Generative AI, LLM, Prompt Engineering}

\maketitle

\section{Introduction}
\label{sec:introduction}

Automated recommendation has become a prominent and universal practical application of machine learning (ML) with potential benefits to both businesses and consumers. As the amount of content available for viewing or purchasing online is ever-increasing, it has become necessary to automate the task of filtering through the vast amount of choices available to humans to avoid information overload. State of the art recommender systems rely on ML algorithms that aim to learn from the patterns of activities and feedback of humans in order to identify and recommend a narrower set of options that are most in-line with these patterns. These patterns manifest in many interactions one has on the world wide web: as implicitly as a view or a click; or explicitly as a rating or purchase. However, this mode of interaction is largely superficial. In most recommendation algorithms implemented in practice there is no direct line of communication between the human and the model. While it can be argued that one may not want to argue with an AI about their preferences when trying to choose a movie to watch or a book to read; the possibility to do so presents an interesting opportunity for a model to learn directly from the user it serves. This is in contrast to the more passive consumption of the user's activity data when building AI recommendation models.

The task of introducing conversation into recommendation has been addressed in the past with varying degrees of success. When Natural Language Processing (NLP) was far less advanced, chatbots such as \cite{thompson2004personalized} were able to tailor the recommendation experience to the user, but were restricted to a handful of response templates. This makes recommendations quick and easy, but with limited scope. However, more recently, techniques that use deep learning have become far more interesting and surprising to engage with.
The encapsulation of Large Language Models (LLM) like ChatGPT into a non-technical, user-friendly interface has further redefined what it means to use and interact with AI systems. This work is interested in ChatGPT specifically because of this ease of access, with the assumption that anyone could go to its interface and request recommendations with little effort. 

In this work we focus on evaluating the direct top-n recommendation potential of the large language model ChatGPT \cite{ouyang2022training}. Recent works \cite{gao2023chat, liu2023chatgpt, kang2023llms} have largely evaluated ChatGPT's recommendation potential on the basis of single inputs and outputs, neglecting its ability to converse. This format often requires the model to choose an option out of a predetermined set that best completes the task; which we argue is not representative of how a user would actually interact with such a system. We choose to structure our study in a way that utilizes the conversational ability of ChatGPT as part of the recommendation process. We further aim to evaluate how ChatGPT performs at recommendation in a more natural setting, i.e. how can it generate pertinent information to the task that has not been provided beforehand. Furthermore, as LLMs have consumed enormous swathes of data in training and some recommendable items may appear more often than others, we are interested in determining whether the model exhibits popularity bias in its recommendation. Our contributions are summarized below:
\begin{itemize}
    \item A pipeline that serves both to rigorously evaluate the recommendation ability of ChatGPT using conversation and be used in practice to refine a set of recommendations for a user.
    \item A prompt engineering approach using iterative feedback, including negative feedback.
    \item An investigation of popularity bias in ChatGPT, and possible ways to mitigate this with pipeline/model parameterization.
\end{itemize}

\section{Related Work}
\label{sec:related_work}

\raggedbottom

\subsection{Prompt Engineering}
Prompt engineering is a new field that has arisen as a result of recent LLMs' in-context reasoning capabilities without fine-tuning. This burgeoning field aims to explore how best to communicate with LLMs when asking them to perform a task in-context. Generally, there have been three predominant means of communication with LLMs through prompting \cite{brown2020gpt3}:
\begin{itemize}
    \item \textbf{Zero-shot}: the model is provided with only instructions and asked to complete a task.
    \item \textbf{Few-shot}: the model is given examples demonstrating a task, and is then asked to repeat this task by generating its own output for a similarly structured question. \textit{One}-Shot specifies that the prompt contains \textit{one} example.
    \item \textbf{Chain-of-Thought prompting (CoT)}: the model is gradually asked to produce intermediate answers before giving the final answer to a multi-step problem \cite{wei2022chain}. The idea is to mimic an intuitive multi-step thought process when working through a reasoning problem.
\end{itemize}
Most other approaches are devised around these archetypes and either vary the amount of information or present the task in a different way.

\subsection{Language Models as Recommenders}

The extensive domain knowledge encapsulated in Large Language Models (LLMs) has recently captured interest for their use in recommendation tasks. Early approaches predominantly concentrated on single input-output tasks. For instance, BERT4Rec \cite{sun2019bert4rec} refines the encoder-only language model, BERT \cite{devlin2018bert}, specifically for sequential recommendations, achieving notable gains over preceding RNN-based benchmarks. LMRecSys \cite{zhang2021language} was among the first methods to explore in-context sequential recommendation with LLMs, showcasing the performance of the encoder-only model and GPT-2 \cite{radford2019language} across a variety of zero shot prompts. Another study \cite{kang2023llms} followed this up with more recent models, with findings that show the benefit of using an LLM for sequential recommendation with fine-tuning. \cite{liu2023chatgpt} underscored the general-purpose recommendation potential of ChatGPT. It emphasized the power of prompt engineering, converting recommendation tasks into natural language tasks, and assessing the model's performance without explicit fine-tuning. \cite{gao2023chat} further evaluated GPT-3 models as augmented language models to interface with existing recommendation systems. They further provided a number of case studies that showcase the potential of ChatGPT for recommendation.

In a recent study \cite{sanner2023large}, researchers explored the potential of LLMs in making recommendations based on both item-based and language-based preferences. Their focus was to compare the efficacy of LLMs with traditional item-based collaborative filtering methods, using a dataset they collected which comprised both types of preferences and their corresponding ratings on recommended items. They showed that LLMs could yield competitive recommendation performance for purely language-based preferences, especially in scenarios resembling a cold-start. \textbf{In contrast, our research focuses primarily on the conversational dynamics of ChatGPT as a recommendation system, emphasizing its interactional capabilities and addressing concerns such as popularity bias through prompt engineering.} Chat-Rec \cite{chatREC2023} is another study close to ours. By translating user profiles and historical data into prompts, it enhances the learning of user preferences and fosters a more direct connection between users and products through in-context learning. In contrast to our work, which emphasizes the iterative feedback and popularity bias mitigation of ChatGPT, the Chat-Rec approach focuses on enhancing traditional recommender systems by converting user data into prompts for LLMs, thereby boosting recommendation interactivity and explainability. Although both approaches address the power of LLMs in the recommendation domain, \textbf{the primary distinction lies in the emphasis of our study on real-time conversational dynamics and Chat-Rec's user-to-prompt data translation for enhancing the recommendation process}.\\

\subsection{Algorithmic Recourse}

Our work has parallels with the concept of \textit{algorithmic recourse} \cite{10.1145/3527848}, which has been defined as "a person's capability to achieve a preferred result from an unchanged model" \cite{Ustun_2019}. By highlighting the individual's ability to acquire a desired outcome from a fixed model, the iterative prompt engineering bears similarities with algorithmic recourse in a way that the user progressively guides the model to a different outcome or prediction. The major difference, however, is that there is no causality involved, but rather a progressive tweaking of the model towards a different outcome. These differences also arise because of the different scopes in algorithmic recourse for the case of a classification output as opposed to a recommendation list output. Another major distinction is that in our context, the algorithmic recourse is driven primarily by the user and not the model. In other words, it is the user who guides the model by providing additional feedback as a follow-up to the initial model output, to guide the model towards an alternative output. Essentially, \textbf{our iterative prompt engineering can be considered as a middle ground between Zero-shot Learning and Algorithmic Recourse}.

\begin{figure*}[ht]
    \centering 
    \includegraphics[width=1\textwidth]{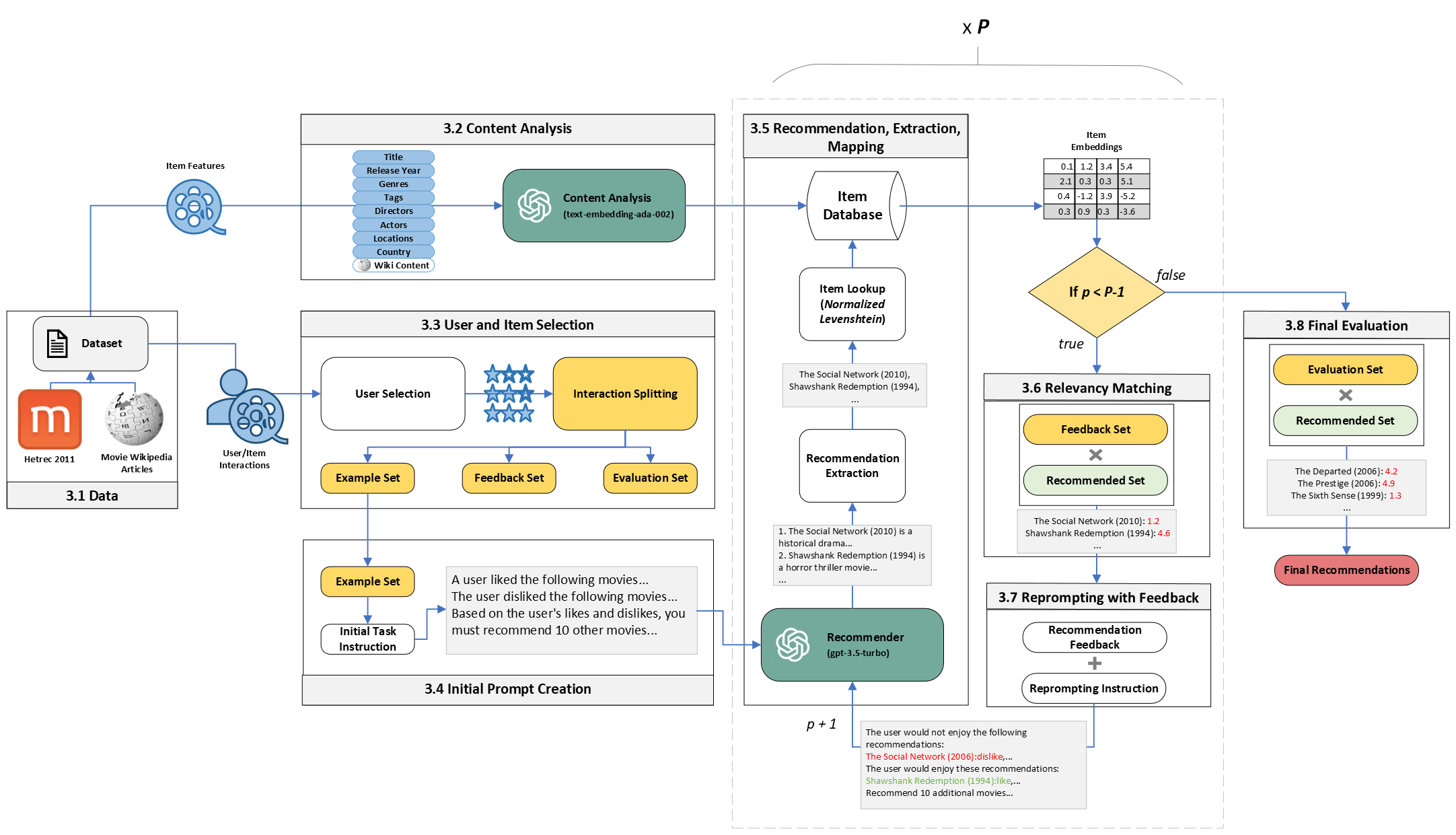}
    \caption{Proposed pipeline for evaluating the effect of conversation in recommendation. P=number of prompts, p=prompt number. Each section corresponds to the section of the same name in the methodology.}
    \label{fig:pipeline}
\end{figure*}

\section{\uppercase{Methodology}}
\label{sec:proposed}

\subsection{Data}

Our study is built around using the HetRec2011 dataset \cite{hetrec} as the ground truth for evaluation. HetRec2011 is an extended version of the MovieLens10M \cite{harper2015movielens} dataset containing additional film information sourced from IMDB \footnote{https://www.imdb.com/} and Rotten Tomatoes \footnote{https://www.rottentomatoes.com/}. Each film contains the following attributes: title, release year, genres, tags, directors, actors, filming locations, and country of origin. To create a comprehensive representation embedding for movie titles in our content analysis, we enriched ChatGPT's information with crawled data from Wikipedia, specifically targeting film-specific articles, successfully obtaining details for 9,722 out of 10,197 movies from the HetRec2011 dataset as of 8/1/2023.

\subsection{Content Analysis}\label{sec:content-analysis}

Our objective is to assess ChatGPT's natural output without pre-providing correct answers, challenging due to its closed-text system without direct ground truth information. We utilize OpenAI's \textit{text-ada-embedding-002} to generate continuous representations for movies in HetRec2011, leveraging their shared GPT-3 foundation. However, gauging exact information learned by LLMs from vast data sources remains complex. While basic descriptors (title, release year, genres) might suffice if a movie has abundant related information in training data, movies without prior data would require an embedding crafted from scratch. Consequently, we generate embeddings for HetRec2011 items, varying the content depth.

Level 1 content contains only the most basic information about a movie, which would be present in smaller datasets like MovieLens100k \cite{herlocker1999algorithmic}. Level 2 incorporates the extra movie attributes from HetRec2011. Level 3 contains additional text scraped from Wikipedia, and level 4 contains level 3 content but with the top 5\% most frequent word-level tokens and stop words removed. We produce this last level based on the traditional IR conception that frequently occurring words are not meaningful for similarity \cite{liu2011web}.

The embeddings come pre-normalized, therefore our only applicable method of computing similarity is cosine similarity. We examine qualitatively how each of these different content levels affect the top-5 most similar items for a given item. An example shows this for the movie `Pineapple Express (2008) in the Appendix Figure \ref{fig:content-comparison}. Based on the description that this movie is a "stoner action comedy film," similar items determined with content level 4 embeddings seem much more appropriate than those found by content level 1 embeddings.

\subsection{User and Item selection}
A sample of 50 users was taken between the 50th and 75th percentile of item interactions, corresponding to a minimum of 122 total interactions. Of these 122 interactions, at least 30 of them must have been dislikes. As long as the constraints are satisfied, the choice of the user is considered arbitrary. It is only necessary that a user has accumulated enough interactions to properly estimate their preferences. This sample was kept small because the ChatGPT output is inherently non-deterministic even when randomness is minimized by setting the $temperature$ parameter to the lowest value. To increase reproducibility and reduce variance in the reported metrics, we instead choose to perform replicate runs for each small user pool to develop a better estimate of the model's performance. Another prohibiting factor resulting in a small sample size is the cost to use ChatGPT's API, which became expensive after testing a multitude of different parameter configurations and conversation lengths.

The user interactions for a user $u \in U$ are split into three random subsets, each serving different purposes throughout the evaluation. We refer to these as the:

\begin{itemize}
    \item \textit{example set}: $\mathbb{E}_u$
    \item \textit{feedback set}: $\mathbb{F}_u$
    \item \textit{evaluation set}: $\mathbb{T}_u$
\end{itemize}

This applies both to positively rated items with ratings $r_i \geq 3$, as well as negatively rated items with a $r_i < 3$ ( for ratings in a scale of 1 to 5). The items in the example set are used in the initial prompt construction to provide ChatGPT with preliminary information on a user's preferences. The feedback set is used solely during the \textit{reprompting} component to help further develop ChatGPT's understanding of the user profile over several iterations. Lastly, the evaluation set is used to evaluate the final set of recommendations. The importance of dividing the interactions amongst these subsets is to separate the information ChatGPT is allowed to learn from versus what it will be tested on. Otherwise, it is probable that the model could simply regurgitate the items given as examples to achieve high performance.

Splits are additionally stratified, ensuring that both the positive and negative rating distributions are similar amongst each of the subsets. This becomes important when trying to determine the relevancy of a recommended item, as ideally, there will be varied positive and negative interactions from the user to estimate how they will feel about an item ChatGPT has recommended.

\subsection{Initial Prompt Creation}

The initial prompt that ChatGPT receives should inform it of its task and optionally provide additional information that helps it to complete this task favorably. For our purposes, we wish to paint an initial picture of a user's tastes and allow ChatGPT to build recommendations based on this. We experiment with three different prompting strategies, as may be seen in Figure \ref{fig:init-prompt} in the Appendix.

Items from the set of examples $\mathbb{E}_u$ are injected into each prompt and appended with an identifier indicating whether an item was liked. Example items are followed by instructional text that requests specific behavior from the model, such as ranking by confidence. Other parameters are injected into the prompt that indicates the number of recommendations requested $k$ and a constraint on the recommendation space by $release\_cutoff$. We specify the latter to avoid making recommendations for items not contained in our dataset \footnote{gpt-3.5-turbo has knowledge up to September 2021 at the time of this research.}. For HetRec2011, the most recent movie was released in 2011.

We recognize that zero-shot is the most likely style that would be used by someone interacting with the system normally. For the sake of comparison, we include prompting styles with examples of the task in one-shot and Chain-of-Thought (CoT) \cite{wei2022chain} format based on findings that prompts displaying higher reasoning have shown an increase in the model's reasoning ability in turn \cite{brown2020gpt3, kang2023llms, liu2023chatgpt}. The CoT format is also motivated by the work of \cite{ye2022unreliability} that shows a marginal increase in performance when including additional explanations. To generate synthetic examples for one-shot and CoT prompts, we randomly sample $|E_u|$ liked and disliked items from all possible items and then construct $k$ recommendations based on the similarity of other items in the data set. The reasoning displayed in CoT also incorporates ranking of results based on sorted collective similarity to the fake liked/disliked items, and aims to demonstrate to ChatGPT how it should approach the problem of making recommendations and sorting them based on confidence. A small detail also found to be impactful; using `\textbackslash n' to separate reasoning steps \cite{fu2022complexity}, has also been included.

\subsection{Recommendation, Extraction, and Mapping}

Once ChatGPT has produced its completion at a given prompting stage, its recommendations must be parsed from the textual output. This output is natural language; thus, some minor errors in formatting, spelling, and grammar are to be expected. In other studies \cite{gao2023chat, liu2023chatgpt}, this has been addressed as a possible issue when extracting and evaluating the recommendations. We noticed little issue with formatting since ChatGPT generally always provides its output as a numbered list of items. Explanations for each item also appear to be automatically provided when the requested number of recommendations is low ($k \leq 10$), and can be turned on or off in the initial instruction prompt. Because the formatting was shown as consistent with numbered lists, it is possible to extract each title using regular expressions and strip extraneous information.

After extracting the titles of each item, they are matched to a generated embedding that has been stored in a database. As noted in \cite{gao2023chat}, ChatGPT does not follow the typical ordering used for film titles, e.g. `\{title\}, \{article\} (\{year\}).' To facilitate better matching, all item titles in HetRec2011 are preemptively re-ordered to match ChatGPT's natural output format of `\{article\} \{title\} (\{year\}).' Furthermore, it is expected that a recommended title does not exactly match the title stored in the database. To avoid problems where a single mistake would cause an exact lookup to fail, the embeddings for the recommended titles are retrieved from the database using Normalized Levenshtein Similarity (NLS)\cite{yujian2007normalized} as a fuzzy lookup approach.

Levenshtein Distance (LD) allows for three single character operations: insertions, deletions, and substitutions; which are used to transform a string $X$ into string $Y$. Typically, the cost of performing each of these three operations is 1. Therefore, the output of LD is merely the number of operations required for the transformation. NLS as defined by Yujian (2007) \cite{yujian2007normalized} constrains this to the interval [0, 1] by weighting the number of changes proportionally to the length of the strings. If we find that a title $X$ is similar enough to a title $Y$ by NLS in the database, we match $X$ with $Y$ and return its embedding. The match similarity threshold is controlled by the parameter $title\_threshold$.

For out-of-dataset items that cannot be matched explicitly or with NLS, we merely exclude these from the metric computation and feedback process. This neither penalizes nor rewards the model for the error. However, as this could skew the evaluation process, we attempt to mitigate the amount of failed matches as much as possible by keeping a short list of the most common unmatched titles. This list is then included in the database, complete with embeddings generated in the same way as was done in the content analysis step described in section 3.2. For fairness of evaluation, there is a need to collect a similar amount of information for them as is available for other films in HetRec2011. From the IMDb non-commercial datasets\footnote{https://developer.imdb.com/non-commercial-datasets/}, we collect the correct release year and genre of the movie, then supplement this with portions of its Wikipedia article as previously described. We collect 363 additional titles that are frequently mismatched.

\subsection{Relevancy Matching}

The simulation pipeline's most important part is estimating whether a given user would respond positively to a recommended item. For a set of recommended items $R_k$ for user $u$, we estimate the rating $\hat{r}_{ui}$ for an item $i \in R_k$ by computing a weighted sum of ratings $r_{uj}$ for item $j \in \mathbb{S}_u$ as:

\begin{equation}
\hat{r}_{ui} = \frac{
\sum_{j \in \mathbb{S}_u} r_{uj} \cdot sim(i, j)  \cdot \id_q(i, j) }
{\sum_{j \in \mathbb{S}_u}  sim(i, j) \cdot \id_q(i, j)} 
\end{equation}

\begin{equation}
    sim (\textbf{u}, \textbf{v}) = \frac{
    \sum_{i=1}^{d} u_i v_i}
    {
    \sqrt{\sum_{i=1}^{d}u^2_i}\sqrt{\sum_{i=1}^{d}v^2_i}
    }
\end{equation}

\begin{equation}
     \id_q(i, j) = \begin{cases} 
      1, & sim(i, j)\geq \epsilon_{q_j} \\
      0, & \text{otherwise} \\
   \end{cases}
\end{equation}

We take inspiration from the rating estimation procedure defined by Sarwar et al. \cite{sarwar2001item}, with the inclusion of an additional indicator function $\id_q$. Where $sim (\textbf{u}, \textbf{v})$ is cosine similarity, and $\epsilon_{q_j}$ is the $q^\text{th}$ quantile of pairwise cosine similarities between item $j$ and all other items in the dataset. If $\hat{r}_{uj} \geq 3$, we accept the recommended item $i$ as a relevant. We use the variable $\mathbb{S}_u$ as a placeholder variable where $\mathbb{S}_u:=\mathbb{F}_u$ when in the reprompting stage, or $\mathbb{S}_u:=\mathbb{T}_u$ when in the final evaluation stage.

This by-item similarity threshold is put in place from a practical standpoint of considering there to be a finite number of items that are realistically comparable to item $j$. For example, specifying that a recommended item must exceed the $99^\text{th}$ quantile of a feedback item's similarities has the practical implication of restricting item interaction to only the top 1\% percent of interactions. For HetRec2011, 1\% of interactions for 10,197 movies means that we consider there to be 101 items that are feasibly comparable to an item of interest. This addition is made because the item representations produced by the text-ada-embedding-002 model are of high dimensionality and thus suffer from a distorted similarity distribution. Making traditional threshold values like 0.5 ineffective for comparison in this context.

\subsection{Reprompting with Feedback}

The reprompting stage involves performing relevancy matching against recommended items, and merely informing ChatGPT which of these recommendations were good or not. We make one small addition to the instructions to ask ChatGPT to avoid making duplicate recommendations. This is to aid in exploring the user's interaction space. In preparing for evaluation, the value for $k$ is substituted by the value for the final number of recommendations $k_f$, and some extra context is added to the next prompt. Both of these prompts can be seen in Figure \ref{fig:reprompt} in the Appendix.

\subsection{Evaluation of Recommendations}

We base our true comparison of different methods only on the last set of $k_f$ recommendations generated only to compare against the evaluation set $\mathbb{T}_u$. Since the entire purpose of a recommendation system is to provide a concise set of relevant items, it would otherwise be impractical to take all $k(p-1)$ + $k_f$ recommendations generated throughout a conversation with $p$ prompts and use this as a measure of quality.

We use several standard metrics to evaluate overall recommendation quality. \textbf{Precision} is used to measure the proportion of ChatGPT's recommendations that would be relevant to the user. \textbf{nDCG} and \textbf{MAP} are used to measure the models ranking performance.

In addition to these typical metrics, we compute several others to derive more information about the model's behavior. The \textbf{Intralist Similarity (ILS)} metric quantifies how similar items in the recommendation list R are to one another \cite{jesse2022intra}. It is commonly computed as the the sum of all pairwise similarities between items $i, j \in R$ divided by the total number of comparisons:

\begin{equation}
    ILS(R) = \frac{\sum_{i \in R}\sum_{j \in R ,i \neq j}sim(i, j)}{(|R|(|R|-1))/2}
\end{equation}

\textbf{Coverage} is a metric that measures how many items are retrievable from the entire catalog of items \cite{kaminskas2016diversity}. This is largely synonymous with Recall, but is modified slightly here to include the quantile threshold $\epsilon_{>_i}$ for an item $i$. This metric captures how many items of user $u$ can be effectively `matched` by a recommended item from ChatGPT, and is computed as:

\begin{equation}
Coverage = \frac{\sum_{i \in R^P, j \in \mathbb{S}_u} match(i, j)}{|\mathbb{S}_u|}
\end{equation}

Where $R^P$ is all recommendations made across all prompts, and $match(i, j)$ is 1 when there exists some item $i \in R^P$ such that $sim(i, j) \geq \epsilon_{q_j}$ for an item $j \in \mathbb{S}_u$, and 0 otherwise. The placeholder variable $\mathbb{S}_u:=\mathbb{F}_u$ (feedback set) or $\mathbb{S}_u:=\mathbb{T}_u$ (evaluation set) depending on the prompting stage. We only count individual occurrences where this is true to show that ChatGPT has approximately recommended an item. 

\textbf{Novelty} is considered to be the inverse of an item's popularity \cite{kaminskas2016diversity}: 1 - $Popularity(i)$, where we compute $Popularity(i)$ across the same pipeline configuration by collecting how many times an item $i$ was recommended to all users $U$ across replicates $\tau$. Formally, we define this as:

\begin{equation}
    Popularity(i) = \frac{\sum_{u \in U} occurs(i, R_u^P)}{|U|\cdot\tau}
\end{equation}

Where $occurs(i, R_u^P)$ is an indicator function that returns 1 if $i \in R^P_u$ for a user $u$ and 0 otherwise. User novelty is computed by summing popularity scores for each item $i \in R^P_u$ and averaging by the total number of recommended items for each user:

\begin{equation}
    Novelty_u = \frac{\sum_{i \in R_u^p} 1- Popularity(i)}{k(p-1) + k_f}
\end{equation}

We define a new metric the \textbf{Unmatched Ratio (UR)} to measure the amount of unmatchable recommendations made for a user $u$. The purpose of this metric is to validate the pipeline by ensuring that other metrics are not being significantly skewed from an inability to match ChatGPT's recommendations to a database item. This is simply defined by:

\begin{equation}
    UR_u = \frac{|\{i \in R^P_u \wedge i \notin I\}|}{k(p-1) + k_f}
\end{equation}

Where $I$ is the set of items in the database, $R^P_u$ is the set of recommendations made across all prompts, $k$ is the number of recommendations made per initial/reprompt, $p$ is the number of prompts, and $k_f$ is the number of recommendations made at the final evaluation prompt.

\subsection{System Parameterization}

We present the total parameterization of the system in the Appendix Table \ref{tab:parameter} for ease of reference.

\section{Experiments}
We conduct experiments that aim to answer the following research questions: \\
\textbf{RQ1:} How does the ability to converse impact recommendation in large language models? \\
\textbf{RQ2:} How do large language models perform at recommendation in their \textit{typical} use-case? (as primarily item-based, top-n recommenders) \\
\textbf{RQ3:} Does Chat-GPT exhibit popularity bias in recommendation? \\

We utilize a Randomized Complete Block Design (RCBD) to account for the variance in the responses induced by different users and different-sized interaction sets. We perform a block on the user and complete 3 full replicates for each block at each level for the independent variables (IV) of interest. Unless otherwise specified, we use an alpha level of $\alpha=0.05$ to determine statistical significance. Provided parametric assumptions hold for an ANOVA, a separate test is ran with each metric as the dependent variable to determine significantly different means. The Kruskal-Wallis test is used as a non-parametric option. Post-hoc tests for significant factors are performed with TukeyHSD after ANOVA, and Dunn's multiple comparison test after Kruskal-Wallis.

We hold the following parameters of the system constant throughout all experiments unless specified:
$k_f=20$, 
$example\_size=10$, 
$eval\_size=0.33$, 
$title\_threshold=0.75$, 
$q=0.99$, 
$temperature=0$, 
$random\_state=22222$.

\subsection{Analyzing the Effect of Embedding Content}

\begin{figure}
    \centering
    \includegraphics[width=0.3\textwidth]{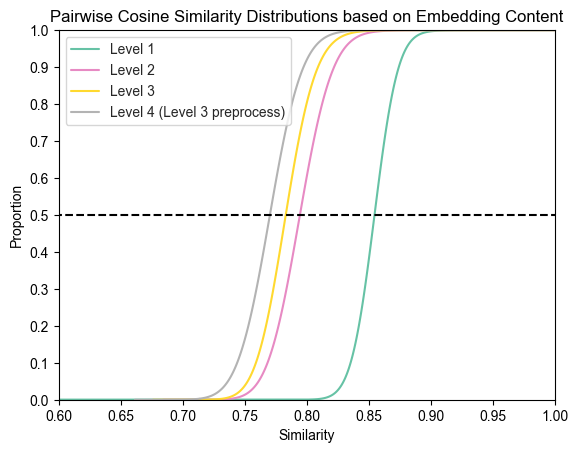}
    \caption{Cumulative distribution functions of item pairwise cosine similarity. Levels are based on the amount of content contained in the sentence embeddings produced by text-davinci-002. The level 4 content level contains the same content as level 3, but with stop words and and the top 5\% most common word-level tokens removed before embedding.}
    \label{fig:cdf-content}
\end{figure}

As a prerequisite to answering our RQs, we first validate the \textit{content} similarity approach by studying how the content used to generate the item embeddings could impact the results. This is because the similarity between items forms the basis of how relevant recommendations are determined. Hence, it is important that their representations allow for a valid comparison.
To get an idea of how the amount of content impacts global similarity, we examine the distributions of pairwise cosine similarity for each content level in Figure \ref{fig:cdf-content}.

It is evident that introducing more content makes similarity more discriminative; however, we seem to approach a limit when incorporating additional information, as indicated by the distribution of levels 2-4. The residual plots of ILS indicated that an ANOVA is not appropriate due to an extreme violation of normality, so we instead utilize the Kruskal-Wallis non-parametric test to determine if there is a significant difference in the levels. P-values for these tests are listed along with the mean metric values for each level in Table \ref{tab:content-means}. We find that for the specified tests, the level of content used in the embeddings is statistically significant at $\alpha=0.01$.

A further analysis for precision, nDCG and AP using TukeyHSD revealed significantly different groupings; and further findings using
Dunn's test with ILS
were consistent with those shown from the initial visualization of the similarity distributions. However, it is interesting to note that even as overall pairwise similarity decreases as the content level increases, the performance also increases. This is in conflict with the initial hypothesis that a greater overall similarity between items makes the process of determining relevant matches less selective. Content levels 2-3 do not produce a significantly different effect from each other, so any of them can be safely chosen to proceed in experimentation. We choose content level 4 embeddings due to a qualitative assessment that they showcase more reasonable similar items. An example of this can be seen in the Appendix in Figure \ref{fig:content-comparison}.

\begin{table}[!t]
\centering
\caption{Mean metric values for different content levels. The model used is ChatGPT with $prompt\_style$=`zero` for $p=1$ prompts. Scores are based on $k=20$, $p=1$ recommendations matched against the evaluation set. *P-value for ILS is determined via Kruskal-Wallis with tie-breaking. We find that content level does have a significant effect on each of the tested metrics at $\alpha=0.01$.}
\label{tab:content-means}
   \aboverulesep=-0.2ex 
   \belowrulesep=0ex
\begin{tabular}{c|cccc}
    \toprule
    Content & Prec. & nDCG & ILS & MAP\\
    \midrule
        1& 0.54 & 0.556 & 0.857 &0.575\\
        2& 0.584 &  0.614 & 0.803 & 0.641\\
        3& 0.581 & 0.619 & 0.804 &0.654 \\
        4& 0.583 & 0.615 & 0.787 &0.649\\
    \midrule
        p-value &$<0.01 $&$<0.01 $& *$<0.01$ &$<0.01$\\
    \bottomrule
    \end{tabular}
\end{table}

\subsection{Analysis of Iterative Feedback}

We aim to answer \textbf{RQ1: How does the ability to converse impact recommendation in language models?}. To accomplish this, we wish to compare different parameterizations of ChatGPT with reprompting against direct recommendation. We vary the following parameters:
\begin{itemize}
    \item $prompt\_style \in \{`zero`, `few`, `CoT`\}$
    \item $k \in \{5, 10\}$
    \item $p \in \{3, 5\}$
\end{itemize}

For comparison against the default direct recommendation task with $k=20$ and $p=1$, we also vary $prompt\_style$. For the few-shot and CoT prompting styles, we only use a single example of $k$ recommendations. All together, there are a total of 15 total settings to compare. Since direct recommendation always only uses the same $k$ and $p$ treatments, these two factors are combined into a new factor $config$ to compare both categories of the system (with and without conversation) in the same tests. We examine interactions through this engineered factor in the post-hoc tests. A two-way ANOVA is used to test for significance in the metrics and we again examine the residuals 
to ensure that the test is valid. ILS is the only result found to violate parametric assumptions. Hence, after an analysis of Kruskall-Wallis with tie-breaking p-values 
we find that all factors are significant, as well as the interaction between $prompt\_style$ and $config$. Post-hoc tests 
later found that the configurations that allow for the most recommendations with $k=10$ were significantly different from other levels for precision, nDCG, and AP. Few-shot and zero-shot are not significantly different, but both perform better than our CoT prompting across the board. Dunn's test results for ILS as DV 
showed that reprompting noticeably increases the similarity between recommendations as the number of prompts increases. CoT prompting significantly decreases ILS, which may indicate that the model fixates on the items we choose as examples.
\begin{figure}[!h]
\centering
     \subfloat[\label{fig:SGD_1_single}]{\includegraphics[width=0.24\textwidth]{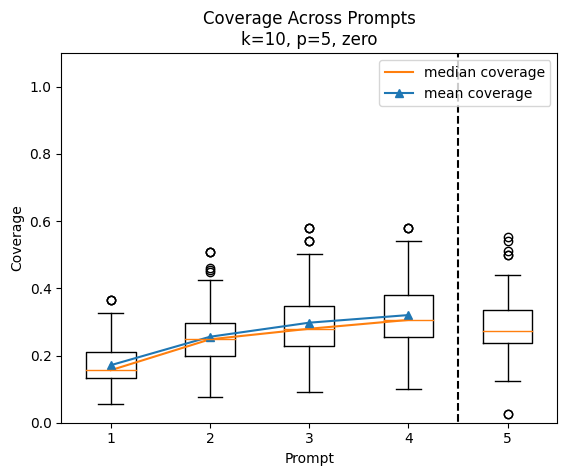}}
     \subfloat[\label{fig:SGD_1_single}]{\includegraphics[width=0.24\textwidth]{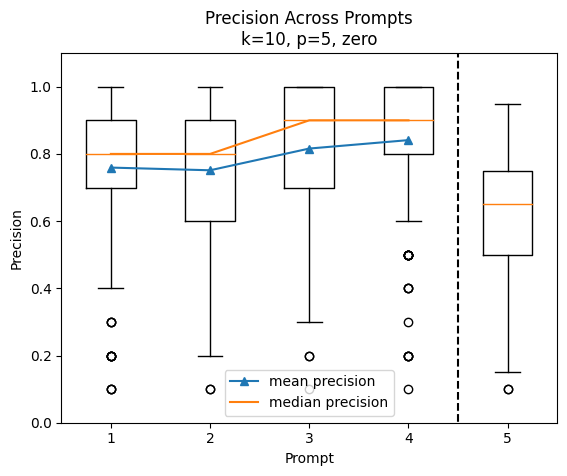}}
     \caption{Coverage and precision distributions for different prompt numbers using best configuration from Table \ref{tab:iterative-comparison}. Plots show that ChatGPT continues to match unique items in the feedback set while further increasing precision.}
     \label{fig:across-prompts}
\end{figure} 

Interactions only show a notable difference between precision configurations, but \textbf{this provides evidence that reprompting is effective in making the model's final recommendations more relevant}. \textbf{With this finding we answer} \textbf{RQ1}. Aggregation of the mean metric values for each possible configuration in Table \ref{tab:iterative-comparison} shows that the configuration $k=10$, $p=5$, and $prompt\_style=`zero`$ appears to be the best parameterization overall based on raw values. We further examine this model in Figure \ref{fig:across-prompts} to see how precision varies with coverage. We see that the model is able to match more relevant items to the user as $p$ increases, which also shows an increase in precision.

\begin{table}[!h]
\centering
   \aboverulesep=-0.1ex 
   \belowrulesep=0ex 
\caption{Mean metric values for different ChatGPT configurations in the pipeline. Scores are based on a final set of $k=20$ recommendations matched against the evaluation set. Best results are colored in each column.}
\label{tab:iterative-comparison}
\begin{tabular}{c|c|c|ccccc}

\toprule
Prompt          & k                   & p & Prec.    & nDCG          & ILS           & MAP           & UR  \\
\midrule

\multirow{5}{*}{Zero} & 20                  & 1 & .586          & .618          & .791          & 0.65          & 6e-4 \\ \cmidrule{2-8}
                      & \multirow{2}{*}{5}  & 3 & .593          & .624          & 0.79          & .647          & 5e-4 \\ 
                      &                     & 5 & .611          & .624          & .788          & .638          & 1e-3 \\ \cmidrule{2-8}
                      
                      & \multirow{2}{*}{10} & 3 & .612          & .653          & .789          & \textcolor{red}{.682}          & 1e-3 \\
                      &                     & 5 & \textcolor{red}{\textbf{.637}} & \textcolor{red}{\textbf{.656}} & .791          & .674          & 1e-3 \\ 
                      \midrule

\multirow{5}{*}{Few}  & 20                  & 1 & .597          & .626          & .792          & .658          & 6e-4 \\ \cmidrule{2-8}
                      & \multirow{2}{*}{5}  & 3 & .616          & .648          & .788          & .676 & 9e-4 \\
                      &                     & 5 & .612          & .62           & .783          & .632          & 2e-3 \\ \cmidrule{2-8}
                      & \multirow{2}{*}{10} & 3 & .629          & .645          & .787          & .664          & 1e-3 \\
                      &                     & 5 & .627          & .645          & .788          & .66           & 5e-3 \\ 
                      \midrule
\multirow{5}{*}{CoT}  & 20                  & 1 & .52           & .527          & \textcolor{red}{\textbf{.743}} & .534          & 1e-2 \\ \cmidrule{2-8}
                      & \multirow{2}{*}{5}  & 3 & .615          & .628          & .784          & .644          & 3e-3 \\
                      &                     & 5 & .602          & .62           & .787          & .638          & 2e-3 \\ \cmidrule{2-8}
                      & \multirow{2}{*}{10} & 3 & .607          & .632          & .786          & .651          & 6e-3 \\
                      &                     & 5 & .625          & .637          & .783          & .653          & 1e-2 \\
                    
\bottomrule
\end{tabular}
\end{table} 

\subsection{Analysis of ChatGPT as a Top-n Recommender}

We perform a comparison between ChatGPT in the pipeline versus baseline models in order to answer \textbf{RQ2: How do language models perform at recommendation in their \textit{typical} use-case? (as primarily item-based, top-k recommenders)}. The two best parameterizations of the pipeline with and without reprompting, as indicated in Table \ref{tab:iterative-comparison}, are selected to represent ChatGPT. We employ four total configurations that utilize NMF as the underlying recommender component. 

\textbf{NMF-item} constructs a set of $k_f$ recommendations for a user $u$ by first building individual lists of $k_f$ unique recommendations for each positive item in the example set, creating a pool $P_u$ of possible recommendations where $|P_u|= k_f * |\mathbb{E}_u|$. This pool is then reduced to the top $k_f$ items with the highest similarity to the items in $\mathbb{E}_u$. \textbf{NMF-user} produces recommendations based on the top $k_f$ items most similar to the user in the user-embedding space. Both models use their own learned embeddings to produce recommended titles. We also vary whether relevancy matching is to be performed using the learned embeddings of NMF, or the GPT-3 embeddings. When using the GPT-3 embeddings, NMF has obviously learned a very different representation for each item because social information has been incorporated. Therefore, what may be an otherwise good recommendation when considering user correspondence is not likely to translate well when evaluated with content-based information only. This inclusion is made regardless to ensure a fair comparison since the only change to the pipeline is how recommendations are produced. We otherwise include the NMF's learned embeddings for relevancy matching to give a more accurate assessment of its performance and the overall effectiveness of the proposed evaluation approach. The results of the evaluation of the NMF model on the interactions in the evaluation split $\mathbb{T}$ can be seen in Table \ref{tab:nmf-actual} in the Appendix.

The mean metric values for each model tested can be seen in Table \ref{tab:model-compare}, along with the p-values for the statistical tests. Post-hoc tests were performed but omitted for space optimization. We find that ChatGPT is significantly better than the Random baseline, which indicates that it is using the knowledge of the user to its advantage. The NMF recommenders evaluated with GPT-3 embeddings perform poorly as expected, but still perform better than random. Interestingly, the NMF models evaluated using their own learned embeddings perform similarly to ChatGPT. This may indicate that ChatGPT with iterative feedback is as effective as a supervised model in the eyes of our evaluation pipeline. However, we refrain from giving too much significance to this claim without testing more model varieties. Based on these findings, we answer \textbf{RQ2} by showcasing that ChatGPT is at the very least preferable to random selection at recommendation.

\begin{table}[!h]
\caption{Comparison of two best pipeline parameterizations (with and without reprompting) against baseline models Non-negative Matrix Factorization (NMF) and Random. Metric values are averaged. *P-value for ILS is determined via Kruskal-Wallis with tie-breaking.}
\label{tab:model-compare}

   \aboverulesep=-0.1ex 
   \belowrulesep=0ex 

\begin{tabular}{l|ccccc}
\toprule
Model                     & Prec. & nDCG & ILS  & MAP  & UR   \\ 
\midrule
(k=10, p=5, zero) & \textcolor{red}{.637}      & \textcolor{red}{.656} & .791 & \textcolor{red}{.674} & 1e-3 \\
(k=20, p=1, few)  & .597      & .626 & .792 & .658 & 6e-4 \\
NMF-item                  & .263      & .262 & .773 & .259 & -    \\
NMF-user                  & .27       & .285 & .77  & .3   & -    \\
Random                    & .243      & .246 & \textcolor{red}{.76}  & .246 & - \\
\midrule
NMF-item (learned) & .626      & .629 & .639 & .634 & -    \\
NMF-user (learned) & .646      & .648 & .637 & .647 & -   \\
\bottomrule
\end{tabular}

\end{table}

\subsection{Exploring Popularity Bias in Recommendation}

Our last experiment tries to answer \textbf{RQ3: Does Chat-GPT exhibit popularity bias in recommendation?} Due to the amount of tests performed, it is apparent that ChatGPT prefers certain recommendations over others; indicating popularity bias. Visualizations of item recommendation frequency are given in Figure \ref{fig:rec-freq} in the Appendix for further reference. Interestingly, we find that the list of most frequently recommended items coincides with the IMDB top 250 movies list \cite{imdb_top_chart}. With this evidence alone \textbf{RQ3} could easily be answered. However, it is more worthwhile to look for a solution to mitigate this popularity bias. To explore this, we choose the following factors and levels:

\begin{itemize}
    \item $prompt\_popular \in \{`no`, `yes`\}$
    \item $temperature \in \{0, 0.5, 1\}$
\end{itemize}

The factor $prompt\_popular$ indicates whether we allow ChatGPT to use popular recommendations. When $prompt\_popular$=`no` we merely add the additional instruction: "Try to recommend movies that are less popular," to the initial prompt and reprompts. The default settings are $prompt\_popular=`yes`$ and $temperature=0$. To determine whether these factors play a role in reducing popularity bias in recommendations, we are mostly concerned with the novelty of the recommendations. However, we performed significance tests for all other metrics as before. Analyzing the residuals of the ANOVAs 
showed similar findings to previous experiments. The P-values for the statistical tests 
indicate that all individual effects and interactions are significant for novelty, but only $prompt\_popular$ is significant for other metrics. A TukeyHSD comparison on $prompt\_popular$ 
showed a significant performance reduction by restricting the recommendation of popular items. 
As $temperature$ increases, the ILS of the recommended items decreases considerably.

\begin{figure}[!h]
\centering
     \includegraphics[width=0.24\textwidth]{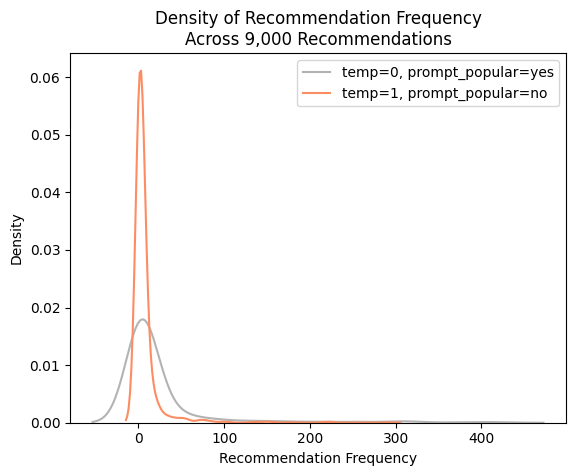}
     \caption{Recommendation frequency by item and and density of two ChatGPT configurations. The model used has parameters $p=5$, $k=10$, $prompt\_style=$`zero.` Prompting with $temperature=1$ and specifying that popular recommendations should be limited ($prompt\_popular$=no) reduces the short-tail of recommendation frequency.}
     \label{fig:bias-reduction}
\end{figure}

The main results of interest are how these factors influence novelty. A TukeyHSD comparison for novelty 
indicated that a high temperature and restricting popular recommendation has a profound effect on recommendation variety. If we wish to maximize novelty, we would choose to use $temperature=1$ and $prompt\_popular$=`no`. The effect of this is better seen in Figure \ref{fig:bias-reduction}, clearly indicating a reduction of the short-tail in item frequency. One caveat to this is that using these options comes at the cost of performance; as indicated by analysis of the other metrics. However, this is a common trade-off associated with traditional approaches to improve variety in recommendation \cite{iaquinta2008introducing, ziegler2005improving}.

\begin{table}[!h]
\caption{Mean metric values for combinations of $temp$ and $prompt\_popular$ (abbr. \textit{pp}) parameters.}
\label{tab:pop-mean}

   \aboverulesep=-0.1ex 
   \belowrulesep=0ex 

\begin{tabular}{c|c|cccccc}
\toprule
Temp          & \textit{pp} & Prec. & nDCG & ILS  & MAP  & Novelty & UR   \\
\midrule
\multirow{2}{*}{0.0} & no             & .43       & .438 & .725 & .448 & .338    & 3e-2 \\ \cmidrule{2-8}
                     & yes            & .637      & \textcolor{red}{.656} & .791 & \textcolor{red}{.674} & .282    & 1e-3 \\ \midrule
\multirow{2}{*}{0.5} & no             & .437      & .444 & .747 & .454 & .407    & 3e-2 \\ \cmidrule{2-8}
                     & yes            & \textcolor{red}{.64} & \textcolor{red}{.656} & .792 & .67  & .288    & 2e-3 \\ \midrule
\multirow{2}{*}{1.0} & no             & .413      & .422 & \textcolor{red}{.712} & .43  & \textcolor{red}{.581}    & 6e-2 \\ \cmidrule{2-8}
                     & yes            & .626      & .643 & .788 & .65  & .357    & 5e-3 \\
                     \bottomrule
\end{tabular}

\end{table}

\section{Conclusion}
\label{sec:conclusion}

We developed an evaluation pipeline centered around ChatGPT, positioning it as a top-n direct conversational recommendation system. While previous studies \cite{liu2023chatgpt}\cite{kang2023llms} have examined ChatGPT's proficiency in selecting optimal recommendations from a candidate pool, our aim was to focus on a more realistic scenario wherein the user does not provide ChatGPT with possible answers ahead of time. Through our investigation, we found that reprompting ChatGPT with feedback during a conversation has a significant impact on recommendation performance over single prompt instances. Furthermore, we showed that ChatGPT is significantly better than both a random and traditional recommender systems, underscoring the utility of its robust domain knowledge in zero-shot recommendation tasks. Finally, we examined ChatGPT's tendency towards popularity bias and proposed strategies to counteract it, paving the way for more novel recommendations.

Despite these informative findings, our system currently has several limitations. First, we are dependent on a relatively large amount of text data in order to produce robust content embeddings. Additionally, we are constrained to relatively old movie releases from our dataset. Future work may require that a more recent dataset be used in order to fully capture how ChatGPT performs on recent information. Another limitation comes from our lack of comparable models to compare with ChatGPT in the pipeline. In a follow-up study, comparing ChatGPT's performance against other LLM and state-of-the-art recommendation algorithms should be further investigated, as was done in other studies \cite{kang2023llms, gao2023chat}.

\bibliographystyle{ACM-Reference-Format}
\bibliography{root}


\appendix
\counterwithin{figure}{section}
\counterwithin{table}{section}
\section{Baseline Recommender Details}

The NMF parameters are found using grid search with a 5\% validation split taken from the non-evaluation items. Training is carried out for 15,000 updates using SGD, with model parameter restoration based on maximum RMSE for the validation set. Using this approach, we find the optimal parameter set as $\lambda=0.05$, $\alpha=1.2$, and $d=50$.

\begin{table}[h!]
\centering
\caption{Average NMF performance as measured on the evaluation set outside of the proposed methodology.}
\label{tab:nmf-actual}

    \begin{tabular}{c|c|c|c|c}
    \hline
         RMSE & Prec.@5 & Recall@5 & MAP@5 & nDCG@5\\
         \hline
         1.181 & 0.796 & 0.052 & 0.672 & 0.553 \\
         \hline
    \end{tabular}
\end{table}
\FloatBarrier

\clearpage
\section{Prompt Examples}

\begin{figure}[!h]
    \centering
    \includegraphics[width=0.78\textwidth]{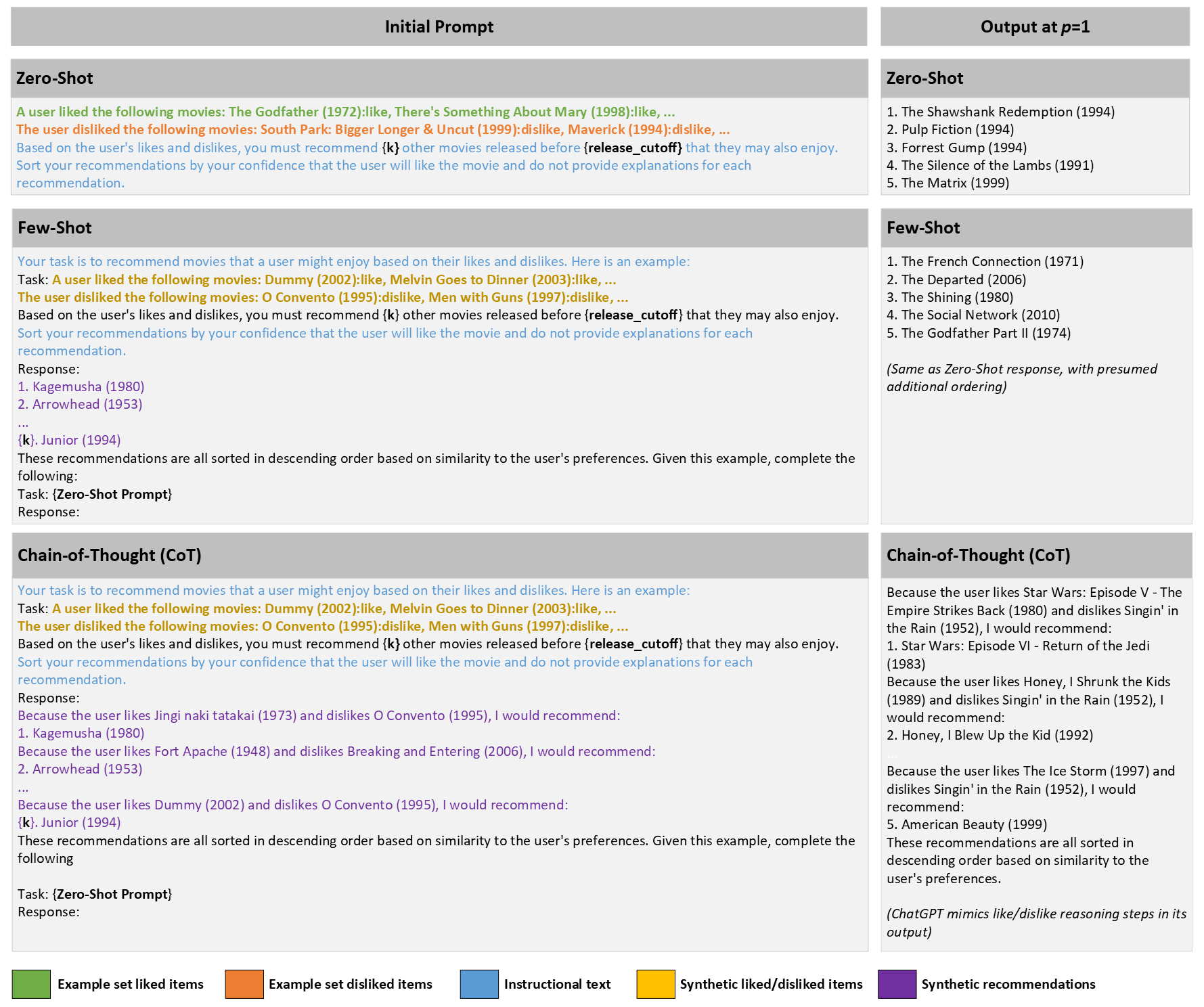}
    \caption{Initial prompt choices. Parameter injection is in bold and contained in `\{-\}` but is represented as only the value in the actual text. }
    \label{fig:init-prompt}
\end{figure}



\begin{figure}[!h]
    \centering
    \includegraphics[width=0.78\textwidth]{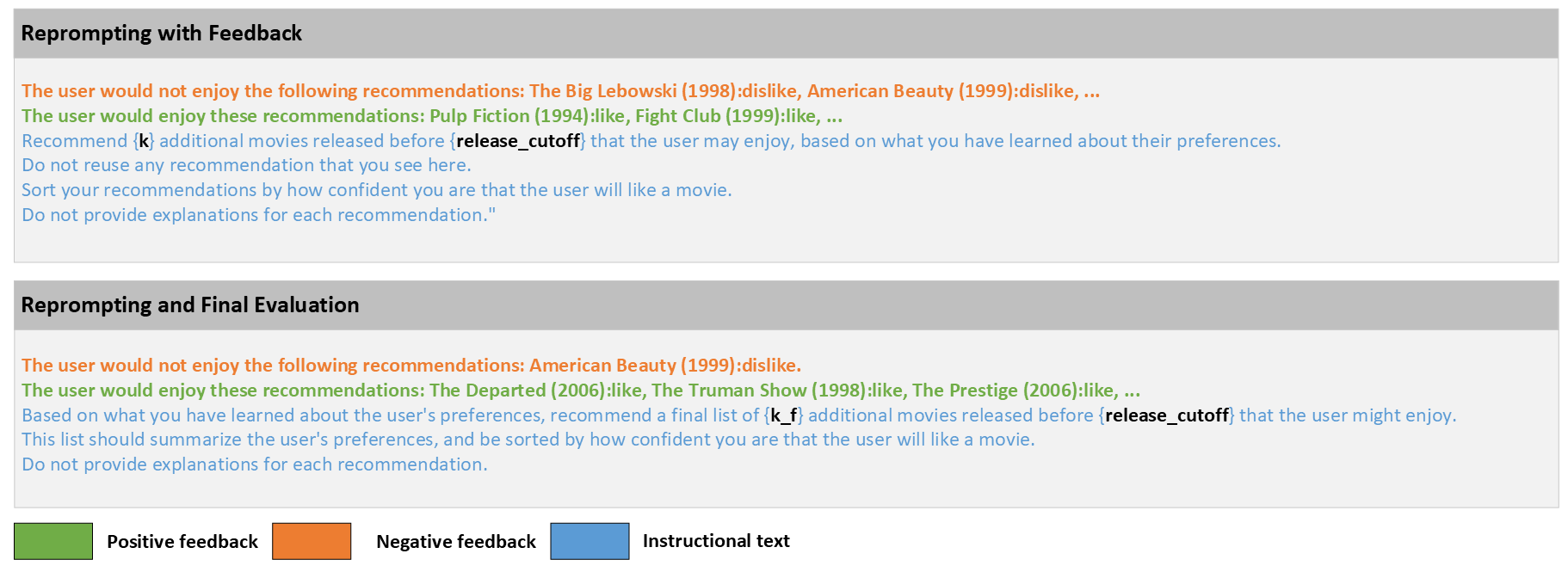}
    \caption{Re-prompting for incorporating feedback mid-conversation and requesting a final recommendation list. Parameter injection is in bold and contained in `\{-\}' but is represented as only the value in the actual text. }
    \label{fig:reprompt}
\end{figure}
\clearpage

\section{Parameter Glossary}

\begin{table}[!h] 
    \caption{Parameterization of System }
    \label{tab:parameter}
    \centering
    \begin{tabular}{|p{0.1\textwidth}|p{0.8\textwidth}|}
    \hline
    $p$ & The number of prompts to which the recommender responds to. Including the initial prompt. \\
    \hline
    $k$ & The number of recommendations to generate at the $p^{th}$ prompt. \\
    \hline
    $k_f$ & The number of recommendations that we request from the recommender at the \textit{final} prompt. These recommendations are ideally a summarized and tuned set of recommendations, after a series of feedback reprompts. \\
    \hline
    $example\_size$ & Determines the number or fraction of interaction tuples $(r_{ui}, i)$ to include in the example set $\mathbb{E}_u$ for a user $u$. \\ \hline
    $eval\_size$ &  Determines the number or fraction of interaction tuples $(r_{ui}, i)$ to include in the evaluation set $\mathbb{T}_u$ for a user $u$. \\ \hline
    $prompt\_style$ & Specifies how the initial prompt will be constructed, namely one of the options in the set: \{`zero`, `few`, `CoT`\}. \\ \hline
    $q$ & The $q^{th}$ quantile of pairwise similarity for an item. Specifies the strictness of weighting in relevancy matching by reserving weighting privilege to only a subset of comparable items for each item. \\ \hline
    $title\_threshold$ &  Similarity threshold in which to accept a recommended title as a match in NLS. \\ \hline
    $model$ & The recommender model component of the simulation pipeline. \\
    \hline
    $temperature$ & Influences the stochasticity of ChatGPT's responses.\\ 
    \hline
    $random\_state$ & Seed shared by all random components of the system. \\
    \hline
    \end{tabular}
\end{table}
\FloatBarrier

\section{Extra Experimental Figures}

\subsection{Content Level Item Comparison}
\begin{figure}[!h]
    \centering
    \includegraphics[width=0.90\textwidth]{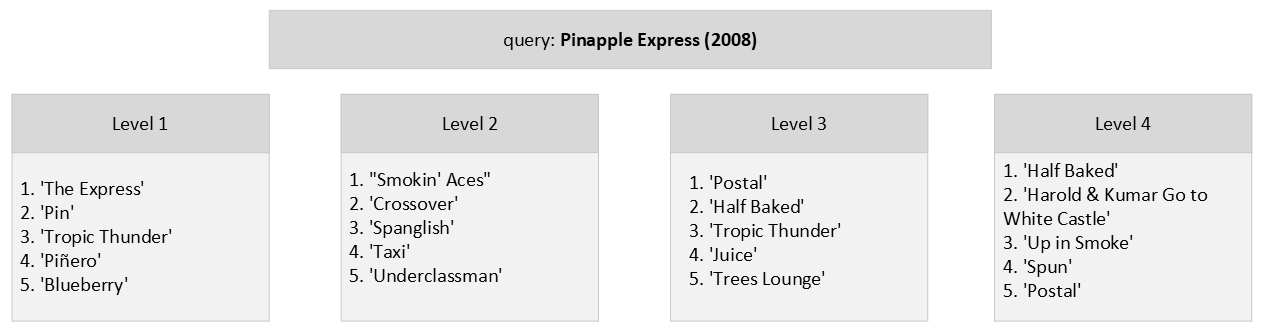}
    \caption{Comparison of the top 5 most similar items to the movie `Pineapple Express (2008)` based on the content level.}
    \label{fig:content-comparison}
\end{figure}

\clearpage
\FloatBarrier
\subsection{Popularity Bias}
\begin{figure}[!h]
    \centering
    \subfloat[\label{fig:SGD_1_single}]{\includegraphics[width=0.55\textwidth]{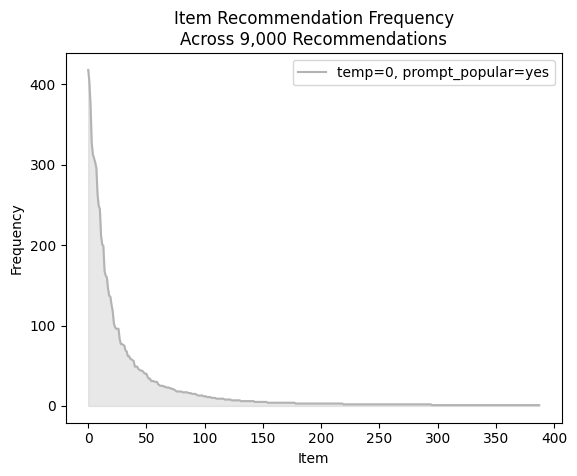}}
     \subfloat[\label{fig:SGD_1_single}]{\includegraphics[width=0.40\textwidth]{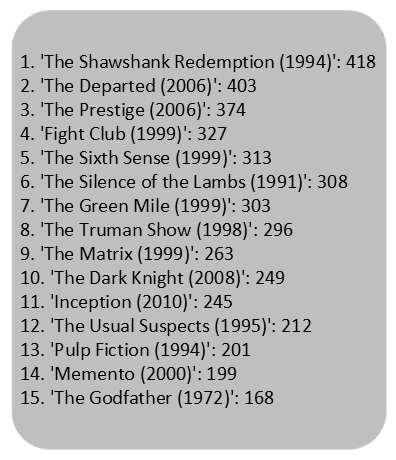}}
    \caption{Frequency of recommendations across 9,000 total recommendation instances.}
    \label{fig:rec-freq}
\end{figure}
\FloatBarrier
    
\end{document}